# Modelling Tritium Production and Release at High-Energy Proton Accelerators

*Dali* Georgobiani [1*], *Thomas* Ginter [2], *Alajos* Makovec [1], *Igor* Rakhno [1], *Matthew* Strait [1], *Igor* Tropin [1]

[1]Fermi Forward Discovery Group, LLC (FermiForward), Fermi National Accelerator Laboratory, Batavia IL 60510, USA
[2]Facility for Rare Isotope Beams, Michigan State University, East Lansing MI 48823, USA

**Abstract.** Tritium is a well-known byproduct of particle accelerator operations. To keep levels of tritium below regulatory limits, tritium production is actively monitored and managed at Fermilab. We plan to study tritium production in the targets, beamline components, and shielding elements of the Fermilab facilities such as NuMI, BNB, and MI-65. To facilitate the analysis, we construct a simple model and use three Monte Carlo radiation codes, FLUKA, MARS, and PHITS, to estimate the amount of tritium produced in these facilities. The analysis could also serve as an intercomparison between these code results related to tritium production. To assess the actual amounts of tritium that would be released from various materials, we employ a semi-empirical diffusion model. The results of this analysis are compared to experimental data whenever possible. This approach also helps to optimize proposed target materials with respect to the tritium production and release.

## 1 Introduction

Tritium is a common byproduct of particle accelerator operations, and its release is actively monitored and controlled at Fermilab to comply with regulatory limits. In this study, we simulate tritium production by modelling a proton beam stopped in a target to estimate tritium yields in various materials commonly used in beamline components in high-energy accelerator environments.

We developed a simplified model to facilitate the analysis and employed three widely used Monte Carlo (MC) radiation transport codes: FLUKA [1, 2], MARS [3, 4, 5], and PHITS [6], to estimate the tritium production rates across various materials of the target assembly and beamline components, including shielding. A model of Fermilab's Neutrinos at the Main Injector (NuMI) target prototype was utilized to represent realistic beam operation conditions. This approach enables a code-to-code intercomparison and provides a valuable tool for optimizing proposed target materials with respect to tritium production. To assess the tritium release from various materials, we employ an analytical diffusion model.

## 2 NuMI Beam Line Concept

Details of the NuMI beam line are shown in **Figure 1**. A beam of 120 GeV/c protons travels through the carrier tunnel and strikes the graphite target in the target hall, producing pions and other particles. Charged pions are focused by toroidal magnets (horns) in the target hall. The decay pipe serves as a long decay path required for pions to decay into neutrinos. The absorber composed of aluminium, steel and concrete stops the remaining hadrons and electromagnetic radiation. Muons are absorbed in the (surrounding) rock, while neutrinos continue through it. The power of the tritium producing particle shower is approximately equally distributed among (i) the target hall, (ii) the decay pipe, and (iii) the absorber.

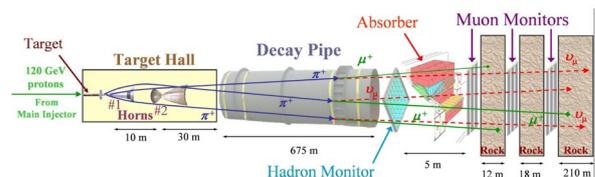

**Figure 1.** NuMI beam line showing the Target Hall, Decay Pipe, Absorber, and Muon Monitors (from [7]).

## 3 Model Input

The NuMI target consists of a series of graphite fins held together by aluminium components (**Figure 2, Figure 3**).

---

* Corresponding author : dgeorgob@fnal.gov

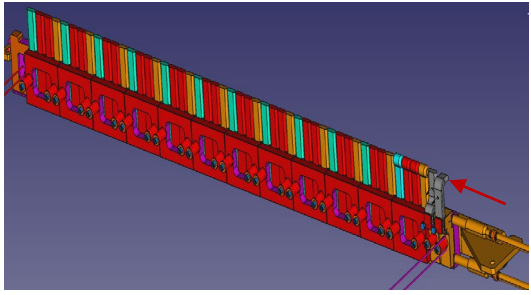

**Figure 2.** Engineering drawing of the NuMI target. The target is constructed from graphite fins (top part) held together by aluminium pieces (bottom part). The red arrow on the right indicates where the beam impinges on the graphite fins.

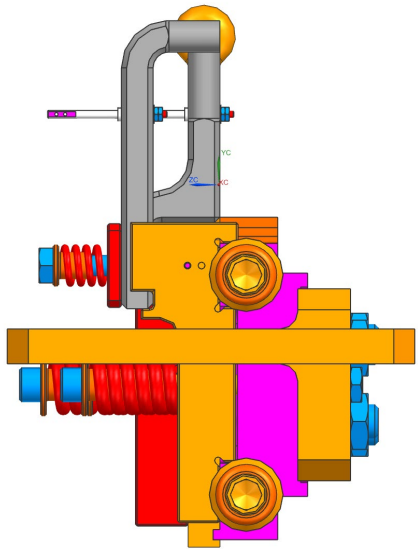

**Figure 3.** Engineering drawing of the NuMI target module, front view.

To estimate tritium production in the target module and in surrounding materials, we used three major Monte Carlo (MC) radiation transport codes: FLUKA, MARS, and PHITS. A simplified version of the target model geometry was constructed, including selected details of its surroundings, such as steel shielding near the target, concrete tunnel walls, and the target water cooling (**Figure 4, Figure 5**).

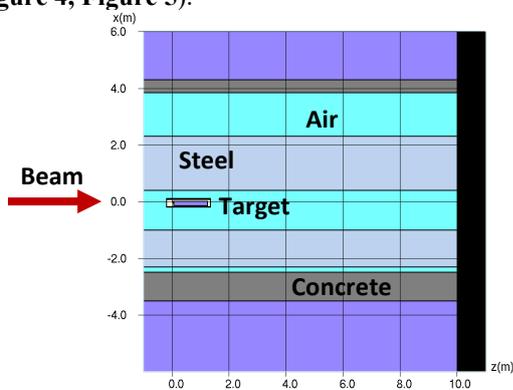

**Figure 4.** Radiation transport model of the NuMI target and its surroundings, elevation view.

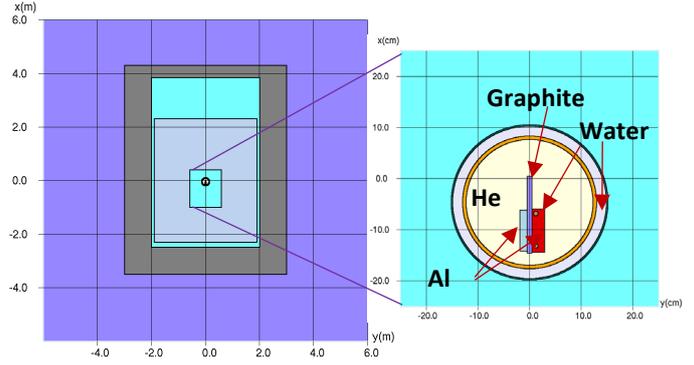

**Figure 5.** Radiation transport model of the NuMI target, cross section (left) and a detailed view (right).

To ensure consistency across the three codes, the model geometry, materials, physics models used in the calculations, energy thresholds and other parameters were made as similar as possible across the three MC codes.

To mimic actual operating conditions, a 120 GeV/c proton beam at 1.1 MW beam power was used, corresponding to an effective beam intensity of 5.25e13 protons per second (pps).

## 4 Tritium Production

Using the three Monte Carlo codes, we calculate tritium production rates in target materials (graphite, aluminium and water) as well as in surrounding shielding (steel and concrete). We also estimate material activation considering a 1-year beam-on scenario. The resulting activation is evaluated immediately after 1 year of irradiation (no cooling) and after 1 year of beam-off (cooling).

### 4.1 Tritium Production Rates

The highest tritium production occurs in the graphite target, the cooling water, and steel shielding (**Figure 6**).

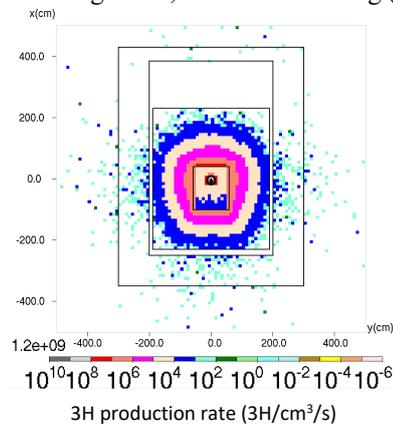

**Figure 6.** Tritium production map in the NuMI target and its surroundings: a cross-sectional view at the location of highest production along the beamline (MARS).

We compare the tritium production rates across the three MC codes (**Table 1**). Only components with significant production rates are shown (very little tritium was produced in aluminium or helium). The ratio between results across the three codes range from 0.4 to 1.5.

**Table 1.** Tritium production rates (3H/cm$^3$/s) in NuMI target and its surroundings, calculated with the three MC codes.

| Region | MARS | FLUKA | PHITS | Volume, cm$^3$ |
|---|---|---|---|---|
| Graphite Target | 5.5E+10 | 3.6E+10 | 7.0E+10 | 8.0E+2 |
| Cooling Water | 2.3E+10 | 4.5E+10 | 6.2E+10 | 1.5E+02 |
| Steel Shielding | 9.2E+05 | 1.5E+06 | 2.1E+06 | 1.7E+08 |
| Concrete Walls | 2.2E+02 | 2.0E+02 | 3.1E+02 | 2.4E+08 |

### 4.2 Activation Due to Tritium

Activation of components after 1 year irradiation is shown in **Table 2** (no cooling) and **Table 3** (1 year of cooling). Similarly to the production rates, the ratio between different MC code results varies between 0.3 to 1.6.

**Table 2.** Total tritium activity in Ci in the NuMI target and beamline component after 1 year of irradiation and no cooling.

| Region | MARS | FLUKA | PHITS |
|---|---|---|---|
| Graphite Target | 64.0 | 41.6 | 79.0 |
| Steel Shielding | 231 | 383 | 520 |
| Concrete Walls | 0.08 | 0.06 | 0.2 |
| Cooling Water | 5.0 | 9.5 | 18.4 |

**Table 3.** Total tritium activity in Ci in the NuMI target and beamline component after 1 year of irradiation and 1 year of cooling.

| Region | MARS | FLUKA | PHITS |
|---|---|---|---|
| Graphite Target | 61.9 | 39.4 | 74.7 |
| Steel Shielding | 219 | 380 | 491 |
| Concrete Walls | 0.07 | 0.06 | 0.2 |
| Cooling Water | 4.7 | 9.5 | 17.4 |

### 4.3 Production Rate Estimates

To understand the differences between the results, we note that the production rate is a product of the high energy (HE) hadron flux, the tritium production cross section, and the target atom density. Production cross sections can be calculated using various methods; one such method is outlined in [8]. This method was used by [9] (see **Figure 7**). For example, comparing tritium production rates in water and steel: the cross section for iron is 10 times higher than for oxygen, whereas the target atom density is roughly twice as high. Therefore, assuming same neutron flux, production rates would be 20 times higher in steel compared to water.

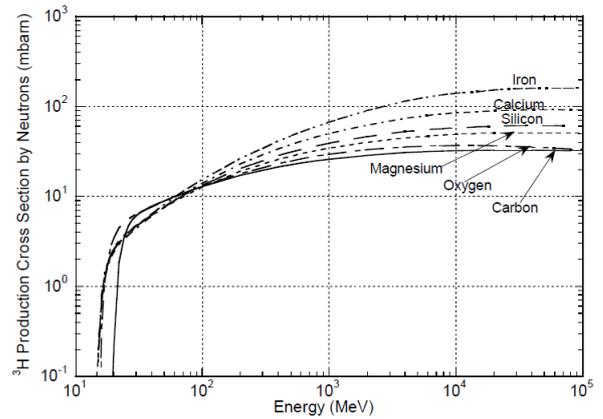

**Figure 7.** Tritium production cross sections due to neutron bombardment of elements commonly found in soil, as a function of neutron energy, from [9].

We calculate the HE hadron fluxes in various materials in the model using the three MC codes, then estimate tritium production rates using the same cross-section value from **Figure 7.** The standard MARS energy threshold of 30 MeV was used for HE hadron fluxes, and the same threshold was applied in FLUKA and PHITS. The HE hadron flux in graphite peaks at 120 GeV because the proton beam strikes the graphite fins. The HE hadron flux in the steel shielding peaks around 100 MeV due to energy loss during transport.

**Table 4.** High energy (HE) hadron flux (particles/cm$^2$/s) in the model components.

| Region | MARS | FLUKA | PHITS |
|---|---|---|---|
| Graphite Target | 1.4E+13 | 1.4E+13 | 1.4E+13 |
| Steel Shielding | 4.3E+08 | 4.3E+08 | 5.2E+08 |
| Concrete Walls | 2.4E+05 | 1.6E+05 | 2.8E+05 |
| Cooling Water | 1.7E+11 | 5.7E+11 | 6.2E+11 |

The HE hadron fluxes are nearly identical in graphite across the three MC codes (**Table 4**), suggesting that differences in tritium production rates are mainly due to variations in production cross sections between codes. For other components, differences in flux also contribute significantly. For example, the MARS result for water is roughly 0.3 times those from FLUKA and PHITS.

Folding in the calculated HE hadron fluxes with the tritium production cross sections from **Figure 7** for various materials, we obtain the production rate estimates and compare them with the ones calculated directly with the MC codes. **Table 5** and **Table 6** show such comparison for the graphite target and the steel shielding, respectively.

**Table 5.** Tritium production rate comparison for the graphite target.

| Production Rate, 3H/cm$^3$/s | MARS | FLUKA | PHITS |
|---|---|---|---|
| MC code | 5.5E+10 | 3.6E+10 | 7.0E+10 |
| Estimate from X-sections | 3.9E+10 | 3.9E+10 | 3.9E+10 |
| Ratio MC/X | 1.4 | 0.9 | 1.9 |

**Table 6.** Tritium production rate comparison for the steel shielding.

| Production Rate, 3H/cm$^3$/s | MARS | FLUKA | PHITS |
|---|---|---|---|
| MC code | 6.8E+05 | 1.1E+06 | 1.5E+06 |
| Estimate from X-sections | 5.5E+05 | 5.1E+05 | 5.9E+05 |
| Ratio MC/X | 1.2 | 2.3 | 2.5 |

## 5 NuMI Tritium Release

Observing tritium behaviour in the actual NuMI setup (**Figure 8**), one can see that at higher beam power, fraction of produced tritium released into air increases significantly. This phenomenon is best explained by the increase in temperature of the steel shielding with higher beam power (**Figure 9**).

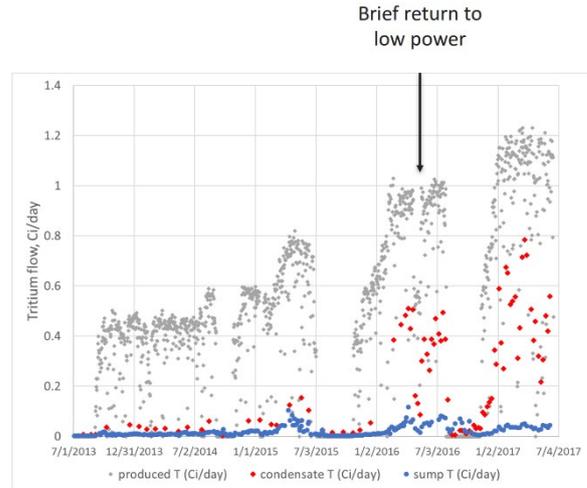

**Figure 8.** Tritium production: Blue – tritium to MINOS sump (to lab water + ponds); Red – tritium collected in condensate and evaporated; Grey – MC results scaled with the number of incident protons (not including absorber) [7].

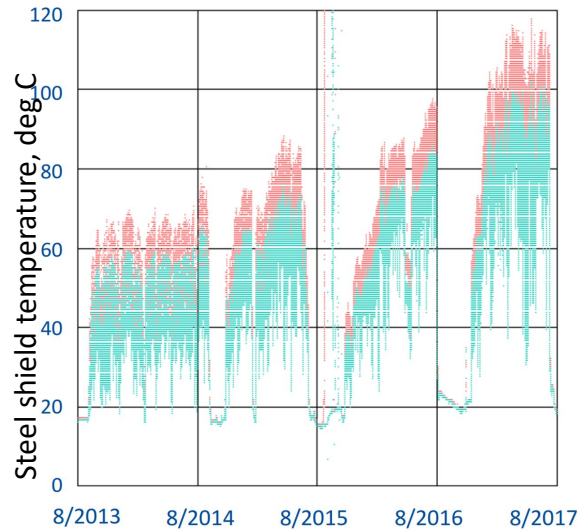

**Figure 9.** Steel shielding temperature increase with increasing beam power.

## 6 Tritium Release

The worst-case scenario assumes that all produced tritium is instantaneously released into the air. In that case, the released tritium would be limited by the production rate alone.

In reality, tritium is released from different materials over varying diffusion times and distances. Diffusion coefficients differ widely between materials and are strongly temperature dependent. For example, the diffusion coefficient for graphite is 8.0E-35 cm$^2$/s at 50°C and 1.2E-11 cm$^2$/s at 1000°C.

## 6.1 Tritium Release Estimates: Formalism

To estimate tritium diffusion from materials, a modified activation equation is used that incorporates both diffusion and ventilation terms:

$$a(t) = a_{sat} \{1-exp(-[\lambda_{3H}+\lambda_{diff1}]t_{irr})\} \{exp(-[\lambda_{3H}+\lambda_{diff2}]t_c)\} \quad (1)$$

Where $a_{sat}$ is saturation activity; $\lambda_{3H}$ is tritium decay constant, or inverse of tritium half-life; $\lambda_{diff1}$ and $\lambda_{diff2}$ are diffusion decay constants for materials at temperatures $T_1$ (during irradiation) and $T_2$ (after cool-down); $t_{irr}$ is irradiation time, $t_c$ is cooling time. Diffusion times are calculated from diffusion lengths L and diffusion coefficients D as $t = L^2/2D$; L is half of the attenuation length for a given material, estimated from MC calculations.

Ventilation is accounted for using a modified saturation activity, $a'_{sat} = a_{sat} * \lambda/(\lambda+r)$, where $\lambda$ is a composite decay constant considering both natural tritium decay and its diffusion, and $r$ is the air exchange rate (number of air changes per unit time).

## 6.2 Diffusion Coefficients, Times, Lengths

Diffusion coefficients for various materials and temperatures are found in literature. To estimate operating temperatures, thermal analysis results from the NuMI target are used (**Figure 10**).

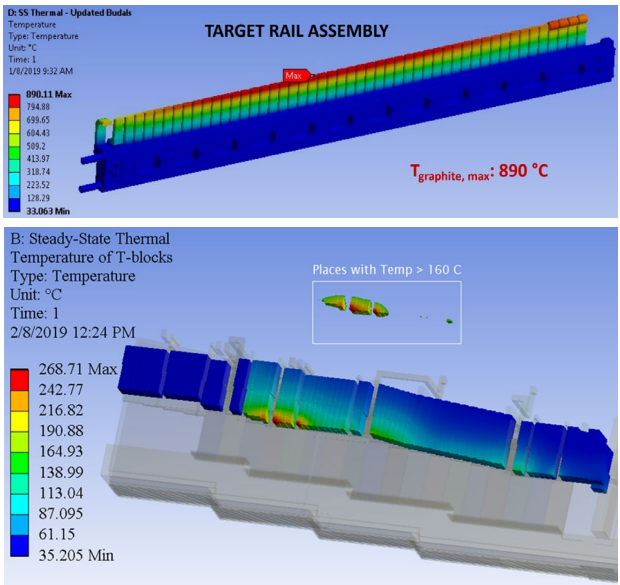

**Figure 10.** Thermal analysis results for the NuMI target assembly (top) and the ambient steel shielding (bottom).

Diffusion effective lengths are conservatively estimated from MC calculations as: (i) half of the attenuation length for a thick (several cm or more) components, or (ii) half of a component thickness if the component is thin compared to the attenuation length.

Diffusion time is estimated as $T_{diff} = L^2/(2D)$, where L is diffusion length.

Diffusion decay constant is $\lambda_{diff} = 1/T_{diff}$; this constant is used, along with the tritium natural decay constant $\lambda_{3H}$, in the modified activation equation (1).

## 6.3 Tritium Release Estimates: Results

Tritium release estimates are based on the production rates from MARS calculations. **Figure 11** shows an example of the diffusion calculation setup.

| Material / Component | H3 production from MC calcs | Temperature T1 (during beam-on) | Diffusion D1 (at T1) | Temperature T2 (during beam-off) | Diffusion D2 (at T2) | Diffusion length | Irradiation time | Cooling time |
|---|---|---|---|---|---|---|---|---|
| | P_3H, 1/s | Beam on | | Beam off | | L, cm | T_irr, h | T_cool, h |
| | | T1, C | D1, cm2/s | T2, C | D2, cm2/s | | | |
| Graphite Target | 4.4E+13 | 1000 | 2.1E-11 | 50 | 8.0E-35 | 0.45 | 8766 | 8766 |
| Iron Shielding | 1.6E+14 | 300 | 2.3E-05 | 50 | 3.2E-06 | 10 | 8766 | 8766 |
| Concrete Walls | 5.1E+10 | 50 | 2.1E-07 | 50 | 2.1E-07 | 12 | 8766 | 8766 |
| **Input** | | | | | | | | |

| Material / Component | Number of 3H after irradiation wo diffusion | Number of 3H after irradiation w diffusion | Fraction diffused | Number of 3H after irradiation & cooling wo diffusion | Number of 3H after irradiation & cooling w diffusion | Fraction diffused |
|---|---|---|---|---|---|---|
| | Ki | Ki_diff | Ri_diff | Kic | Kic_diff | Ric_diff |
| Graphite Target | 1.4E+21 | 1.4E+21 | 2.2E-03 | 1.3E+21 | 1.3E+21 | 2.2E-03 |
| Iron Shielding | 4.8E+21 | 4.9E+20 | 9.0E-01 | 4.5E+21 | 1.1E+20 | 9.7E-01 |
| Concrete Walls | 1.6E+18 | 1.5E+18 | 3.1E-02 | 1.5E+18 | 1.3E+18 | 9.1E-02 |
| **Results** | | | | | | |

**Figure 11.** Screenshots of the tritium release calculation setup: input values (top table) and results (bottom table).

After 1 year of irradiation, 90% of the tritium produced in iron shielding is diffused from steel (see the yellow cell in the lower table of **Figure 11**). After 1 year of irradiation and 1 year of cooling, this number increases to 97% (red cell in the lower table of **Figure 11**).

At a lower beam power, iron shield temperatures are lower; therefore, diffusion coefficients are lower, and less tritium is diffused from materials. For example, under room temperature in the same setup, only about 50% of the produced tritium releases from the steel shielding.

It should be noted that while numerous diffusion coefficient values are available in the literature, they can vary significantly across different sources.

## 7 Summary

We study tritium production and release in Fermilab's NuMI target components. Tritium production is calculated using three major Monte Carlo radiation codes: FLUKA, MARS, and PHITS. This analysis also serves as an intercomparison of these codes with respect to the tritium production. The resulting production rates are generally consistent across the three codes, with minor differences explained. To estimate tritium diffusion from various materials, we apply a modified activation equation that incorporates diffusion and ventilation effects. The results of our analysis offer a detailed explanation for the observed increases in tritium release correlated with increased NuMI beam power in recent years. This methodology will also allow us to make predictions for the future tritium releases at various Fermilab facilities. Finally, this approach helps to optimize the proposed target and beamline materials with respect to their tritium production and diffusion properties.

DG would like to express her gratitude to the SATIF-16 organizers, the SOC and LOC members, and the INFN-LNF management and administrative personnel for the perfectly organized workshop and for their kind hospitality.

This work was produced by Fermi Research Alliance, LLC under Contract No. DE-AC02-07CH11359 with the U.S. Department of Energy, Office of Science, Office of High Energy Physics. Publisher acknowledges the U.S. Government license to provide public access under the [DOE Public Access Plan](#)